\documentclass[a4paper, preprint]{elsarticle}

\usepackage[symbol*]{footmisc} 
\usepackage{hyperref}
\usepackage{amsfonts,amssymb,amsmath,mathtools} 
\usepackage{graphicx}
\usepackage{xcolor}

\usepackage{caption}
\usepackage{subfig}
\usepackage[absolute,overlay]{textpos}

\newcommand {\proof} {\par\textit{Proof}. \ignorespaces}
\newcommand {\eproof}
      {\null\hfill{\large$\Box$}}

\newcommand{\vx}{\mathbf{x}}

\usepackage{multicol}
\newcommand{\fd}[2]{\frac {d#1}{d#2}} 

\newcommand{\eexp}[1]{\text{e}^{#1}}          
\DeclareMathOperator{\Lagr}{\mathcal{L}}      
\DeclareMathOperator{\Act}{S}                 
\newcommand{\Rreal}{\ensuremath{\mathbb{R}}}  
\DeclareMathOperator{\Zint}{\mathbb{Z}}       

\makeatletter                                 
\providecommand{\doi}[1]{
  \begingroup                                 
    \let\bibinfo\@secondoftwo
    \urlstyle{rm}%
    \href{http://dx.doi.org/#1}{%
      doi:\discretionary{}{}{}%
      \nolinkurl{#1}%
    }%
  \endgroup
}
\makeatother

\DefineFNsymbols*{lamport}{*{\textsuperscript{\textdagger}}\ddagger\S\P\|%
{**}{\dagger\dagger}{\ddagger\ddagger}}

\setfnsymbol{lamport}

\makeatletter                
\def\ps@pprintTitle{
 \let\@oddhead\@empty
 \let\@evenhead\@empty
 \def\@oddfoot{}%
 \let\@evenfoot\@oddfoot}
\makeatother

\begin{document}

\begin{frontmatter}

\title{On the efficient numerical solution of lattice systems with low-order couplings}

\renewcommand{\thefootnote}{\fnsymbol{footnote}}

\author[e]{A.~Ammon}
\ead{andreas.ammon@desy.de}

\author[b]{A.~Genz}
\ead{genz@math.wsu.edu}

\author[c]{T.~Hartung}
\ead{tobias.hartung@kcl.ac.uk}

\author[a]{K.~Jansen}
\ead{karl.jansen@desy.de}

\author[d]{H.~Le\"ovey}
\ead{leovey@math.hu-berlin.de}

\author[a]{J.~Volmer}
\ead{julia.volmer@desy.de}

\address[a]{NIC, DESY Zeuthen, Platanenallee 6, D-15738 Zeuthen, Germany}
\address[b]{Department of Mathematics, Washington State University, Pullman, WA 99164-3113 USA}
\address[c]{Department of Mathematics, King's College London, Strand, London WC2R 2LS, United Kingdom }
\address[d]{Institut f\"ur Mathematik, Humboldt-Universit\"at zu Berlin, Unter den Linden 6, D-10099 Berlin}
\address[e]{OAKLABS GmbH, Neuendorfstr. 20b, 16761 Hennigsdorf, Germany}


\begin{textblock}{15}(29,2)
\setlength{\parindent}{0cm}
DESY 15-033
\end{textblock}

\begin{abstract}
We apply the Quasi Monte Carlo (QMC) and recursive numerical integration
methods to evaluate the Euclidean, discretized time 
path-integral for the quantum mechanical anharmonic oscillator and a topological
quantum mechanical rotor model. For the anharmonic oscillator both methods
outperform standard Markov Chain Monte Carlo methods and show a significantly
improved error scaling.
For the quantum mechanical rotor we could, however, not find a successful way
employing QMC. On the other hand, the recursive
numerical integration method works extremely well for this model and shows
an at least exponentially fast error scaling.
\end{abstract}


\end{frontmatter}

\section{Introduction}\label{sec:Intro}

Markov Chain Monte Carlo (MCMC) is the method of choice for simulations of quantum field theories or systems in statistical physics. The advantage of MCMC is that it can be applied very generally to many physical models. It allows to compute expectation values of physical observables $\langle O\rangle$ with an error $\Delta$ which scales only as $\Delta \propto 1/\sqrt{N}$, however, where $N$ is the number of samples. This error scaling law leads to a very large numerical effort if another significant digit in the accuracy of an observable is needed.  

In quantum field theory, in particular quantum chromodynamics (QCD) -
our theory of the strong interaction between quarks and gluons - very significant progress has been achieved in the last years through improvements of the MCMC methods used; see, e.g., ref. \cite{Luscher:2010ae} for an overview. But, even though lattice QCD simulations of the theory could be accelerated substantially, computations typically run several months or even years on state of the art supercomputers. In addition, none of those improvements have changed the error scaling of $1/\sqrt{N}$. It will therefore be very demanding to obtain a significant improvement of the accuracy of physical observables in this field. 

On the other hand, it is known that Quasi Monte Carlo (QMC) \cite{DP10,Dick_Kuo_Sloan_Acta13} or recursive numerical integration \cite{Genz86,Hayter06} methods show a much improved error scaling. For QMC methods this error scaling reads $\Delta \propto 1/N^\alpha$ where $\alpha$ can reach values of $\alpha=1$ or even larger. For recursive numerical integration methods the error scaling can be even better and, in some cases, it is even faster than exponential, see also the discussion below. 

Clearly, such QMC and recursive numerical integration methods could, thus, lead to a much enhanced accuracy of simulations. However, these methods have not been tried for generic quantum field theories, so far, and neither their applicability nor whether they lead to an improved error scaling is clear. 

In \cite{Jansen:2012gj,Jansen:2013jpa,Ammon:2013yka} we initiated a test of QMC methods for the harmonic and anharmonic quantum mechanical oscillator discretized on a Euclidean time lattice; see the next section for an introduction to these systems. The results of these investigations have been very promising. For the harmonic oscillator, which is a Gaussian system, we found an error scaling of $\Delta \propto 1/N$ which is optimal. Adding a non-Gaussian term in case of the anharmonic oscillator, we found $\alpha\approx 0.75$ which is not optimal but significantly better than the error scaling of MCMC methods. However, it needs to be mentioned that for certain system sizes, that is, large Euclidean times, the QMC method did not work particularly well for the anharmonic oscillator and no error scaling improvement could be established; cf., \cite{Jansen:2013jpa} for details.

However, QMC methods can also solve another problem: 
when using MCMC for the anharmonic oscillator, only samples in the 
vicinity of one minimum of the action are drawn and the probability to 
jump to another minimum is very small. This leads to a very large, in 
fact exponential, autocorrelation time. 
As demonstrated in refs.~\cite{Jansen:2012gj,Jansen:2013jpa,Ammon:2013yka} 
with QMC using the harmonic approximation this problem is completely
overcome. This is not surprising since QMC avoids long autocorrelation 
times essentially by construction. We stress that, with the new 
developments described in section \ref{ssec:ExperimentsAnharmonicOszi}, the iterated numerical 
integration method allows to solve the anharmonic oscillator without 
any problem of autocorrelations, as well.

In this paper, we want to extend the work of refs.~\cite{Jansen:2012gj,Jansen:2013jpa,Ammon:2013yka} in two directions. One direction is to apply recursive numerical integration methods for the anharmonic oscillator. As we will see below, this method does not suffer from the shortcomings of the QMC method for large Euclidean times. The other direction is the investigation of a topological quantum mechanical model, the quantum rotor. This model has some characteristic features that are also present for non-linear $\sigma$-models or gauge theories which are essential and most important models for describing elementary particle interactions; see, e.g., refs. \cite{Rothe:1992nt,Montvay:1994cy,Gattringer:2010zz} for introductions to lattice field theories applied to particle theory. However, since the quantum rotor is a much simpler 1-dimensional quantum mechanical model, it is easier to treat numerically. In this way, it becomes possible to test in detail, whether QMC or recursive numerical integration methods work for such a model. Clearly, in case the application of one of the methods fails, it will become almost impossible to proceed with higher dimensional gauge theories. As we will discuss below, so far we have not been able to apply QMC successfully to the quantum rotor. On the other hand, recursive numerical integration methods turn out to be highly successful and show an extremely good error scaling behavior.

Another aspect investigated here is the rapid increase of the autocorrelation time for the quantum mechanical rotor when the continuum limit is approached. For the sake of illustrating this generic problem of MCMC methods, we will perform a calculation of the quantum rotor with the Metropolis algorithm. QMC and recursive numerical integration methods do not suffer from this problem and are, hence, also very advantageous in this respect. Since for the quantum mechanical rotor a particular MCMC method which also avoids the problem of very large autocorrelation times, the cluster algorithm, can be applied, we compare the recursive numerical integration method to this optimal MCMC algorithm to see whether a gain can still be found. 

The paper is organized as follows. In the next section we introduce
the models which are investigated throughout the paper. In section 3 we shortly summarize our (failed) attempts to solve the quantum rotor with QMC methods. In section 4 we introduce the recursive numerical integration method and discuss its theoretical basis. Finally, in section 5 we present our results and conclude in section 6.

\section{Lattice systems: 1-dimensional models}\label{sec:1D-Model}
In this paper we investigate two different quantum mechanical models
in Euclidean time\footnote{Euclidean time is common practice in the
  standard lattice approach.}. In order to evaluate the models
numerically we will discretize time and solve the resulting high
dimensional integrals numerically. In this section we start with some
general considerations about calculating observables of lattice
systems before we describe the two models we investigate, the topological oscillator and the anharmonic oscillator.

\paragraph{Calculating observables in discretized time}
The coordinate $x(t) \in \Rreal$ describes the trajectory of a
particle in time. Classically, this path of a particle propagating
from $x_0 = x(0)$ to $x_1 = x(T)$ during a time period $T$ is
determined by the minimum of the action $\Act(x)$ (with respect to
$x(t)$),
\begin{align*}
  \Act(x) = \int_{0}^{T} \Lagr(x,t) dt  \qquad \in \Rreal,
\end{align*}
with the Lagrangian $\Lagr(x,t)$ containing all necessary information
about the model.

In order to obtain a numerically tractable and positive action while
keeping the quantum mechanical observables real we have to define the
trajectory $x(t)$ on a Euclidean, equidistant time lattice with
lattice spacing $a$. Corresponding to this, we replace the following
quantities by their discretized counterparts:
\begin{align}
  \begin{aligned}
    t &\rightarrow t_i := i \cdot a, &i \in \{0,1, 2, ..., d-1\},\\
    x(t) &\rightarrow x_i := x(t_i), &x_i \in \Rreal,
    \label{equ:disc_t_x}
  \end{aligned}
\end{align}
with $T = a \cdot d$, such that
\begin{align}
  \int_0^T dt \rightarrow a \sum_{i=0}^{d-1}, \qquad 
  \fd{x}{t} \rightarrow \nabla x_i = \frac{1}{a} (x_{i+1} - x_i).
  \label{equ:disc_int_deri}
\end{align}
We remark that the choice of the discretization of the derivative is
not unique and alternative discretizations can lead to different error
expansions of observables in terms of the lattice spacing in the
continuum limit, $a \rightarrow 0$ and $T \rightarrow \infty$. We will use cyclic
boundary conditions $x_{d} :=x_{d(\bmod d)} = x_0$ throughout this paper.

In a quantum mechanical system not only the classical path contributes to a given observable, but all possible paths have to be taken into account. Following Feynman's description, the quantum mechanical system is defined by the path integral 
\begin{align}
  \int_{D^d} \eexp{-\Act[x]} dx , \label{equ:pint}
\end{align}
where $D$ is a domain in $\Rreal$. In \eqref{equ:pint} the
transformation to Euclidean time has been performed already. For a
time discretized quantum mechanical system \eqref{equ:pint} is
well-defined, although it may have a high dimension $d$, which could
be 1000 or larger.

The expectation value $\langle O[x] \rangle$ of an observable $O[x] = O(x_0,x_2,\dots,x_{d-1})$ of a quantum mechanical model with a discretized action $S[x]$ can now be calculated using the path integral formalism
\begin{align}
  \langle O[x] \rangle = \frac
  {\int_{D^d} O[x] \, \eexp{-\Act[x]} dx} 
  {\int_{D^d} \eexp{-\Act[x]} dx} . \label{equ:obs}
\end{align}

We note that the formalism for treating quantum
mechanical systems, as it is sketched out above, can be generalized to
quantum field theories. Such quantum field theories are the actual
basis for the theoretical investigation of elementary particle
interactions.

\paragraph{Topological oscillator}
As indicated previously, the first model we are going to consider in
this article is the topological oscillator or quantum rotor which
describes a particle with mass $M_0$ moving on a circle with radius
$R$, and correspondingly, has a moment of inertia of $I = M_0 R^2$. We
investigate this particular model because it goes beyond the classical
quantum mechanical oscillator (described later on) and already shows some characteristic features of non-linear $\sigma$-models and gauge theories which are of prime importance in particle physics. 

The free coordinate of the system is the angle $\phi \in [-\pi, \pi)$,
describing the position of the particle on a circle with radius $R$
around the origin. The system is described by the action
\begin{align*}
  S(\phi) = \int_{0}^{T} \frac{I}{2}  \left( \fd{\phi}{t} \right)^2 dt,
\end{align*}
and is obtained from the action of a free particle moving in two dimensions,
\begin{align}
  \Act(x) = \int_{0}^{T} \frac{M_0}{2} \left( \dot{x}(t)^2 + \dot{y}(t)^2 \right) dt \quad (x,y) \in \Rreal^2,
\end{align}
and the transformation $x(t) = R \cos(\phi(t))$ and $y(t) = R \sin(\phi(t)) $ respectively.
The corresponding discretized action reads
\begin{align}
  S[\phi] = \frac{I}{a} \sum_{i=0}^{d-1} \left( 1 - \cos \left(\phi_{i+1}
    - \phi_i  \right) \right),
\label{equ:rotoraction}
\end{align}
where we use the cosine to describe the kinetic part of the action, as
$\frac{1}{a^2}(1 - \cos(\phi_{i+1} - \phi_i)) = \frac{1}{2} \left(
  \nabla\phi_i \right)^2  + O( (\nabla \phi_i)^4 ) $ and hence corresponds to leading order to the naive lattice derivative (as in \eqref{equ:disc_int_deri}, right).
In the following we will, if not stated otherwise, set the lattice spacing 
to $a=1$. This means, in particular, that the continuum limit 
is reached\footnote{In principle, the physical extent of time lattice 
$T=d\cdot a$ should be kept constant and hence the continuum limit $a\rightarrow 0$ 
requires $d\rightarrow\infty$.} by $T=d\cdot a \rightarrow \infty$.

One characteristic quantity of the quantum mechanical rotor is 
the topological charge of the system. 
It describes the number of complete revolutions the rotor performs
during a time period $T$
\begin{align*}
  Q(\phi) = \frac{1}{2\pi} \int_{0}^{T} \left( \fd{\phi}{t} 
  \right)dt \qquad\in \Zint.
\end{align*}
We use the discretized version
\begin{align*}
Q[\phi] = \frac{1}{2\pi} \sum_{i=0}^{d-1}  \left( 
    \phi_{i+1} - \phi_i \right) \bmod [-\pi, \pi).
\end{align*}
As an observable, 
we will investigate the topological susceptibility
\begin{align*}
  \chi_t = \frac{\langle Q^2[\phi] \rangle}{T}
  &\stackrel{T \rightarrow \infty}{\longrightarrow} \frac{1}{4\pi^2 I},
\end{align*}
where $\langle Q^2[\phi] \rangle$ is calculated according to \eqref{equ:obs}.  

Other important observables of the system are the energy gaps which
can be extracted from Euclidean correlation functions $\Gamma(j)$,
\begin{align}
  \Gamma(j) = \frac{1}{d} \sum_{i=0}^{d-1} \phi_i \cdot \phi_{i+j},
  \qquad j \in \{0,1,\dots,d/2\}. 
  \label{equ:topo_corr}
\end{align}
It measures the correlation of angles at different lattice sites 
separated by a distance $j$. 
This correlation function has an exponential decay rate with the 
distance $j$,
\begin{align}
  \Gamma(j) \,\tilde{\propto}\, \eexp{-j \cdot \Delta E},
  \qquad{j \gg 1}.
  \label{equ:topo_corrE}
\end{align}
From this the energy gap $\Delta E$ between the ground state and the
first excited state and therefore also the correlation length $\xi$ can
be determined,
\begin{align}
  \xi = \frac{1}{\Delta E} \stackrel{T \rightarrow \infty}{\longrightarrow} 2 I.
\end{align}
Other energy levels can be computed in a similar way.

Through the energy gap, a connection to the topological susceptibility 
can be established, as well; namely,
\begin{align*}
  \chi_t = \frac{\langle Q^2 \rangle}{T}
  &= \frac{1}{2\pi^2} \Delta E.
\end{align*}
This, however, only holds in the continuum limit $a\rightarrow 0$ and
$T\rightarrow \infty$.

\paragraph{Anharmonic oscillator}
The second quantum mechanical system considered in this article is   
the harmonic oscillator which describes a particle with mass $M_0$ moving
along a path $x \in \Rreal$ in
a potential proportional to $x^2$. Adding a $x^4$ term
to the potential, 
the system is called the anharmonic oscillator.
The Lagrangian is given by
\begin{align*}
  \Lagr(x,t) = \frac{M_0}{2} \left( \fd{x}{t} \right)^2 +
  \frac{\mu^2}{2} x^2 + \lambda x^4
\end{align*}
where $\mu^2, \lambda \in \Rreal$.
In order to keep the action bounded from below, the coupling 
$\lambda$ has to be chosen positive. 
With $\lambda>0$ present, the constant $\mu^2$ can be chosen arbitrarily 
and for $\mu^2 >0$ one finds a distorted harmonic potential while for 
$\mu^2<0$ a double well potential appears. 

Using the discretization scheme mentioned in \eqref{equ:disc_t_x} and
\eqref{equ:disc_int_deri}, the lattice action for the
anharmonic oscillator becomes
\begin{align*}
  \Act[x] = a \sum_{i=0}^{d-1} \frac{M_0}{2} (\nabla x_i)^2 +
  \frac{\mu^2}{2} x_i^2 + \lambda x_i^4.
\end{align*}
One type of characteristic observables of this system are powers of the position $x$
of the particle, i.e., $\langle x\rangle$, $\langle x^2\rangle$, $\langle x^4\rangle$, etc.

Another type of observable is again the energy gap which can be
extracted from the correlator $\Gamma(j)$, as already described in
eq.~\eqref{equ:topo_corr} and \eqref{equ:topo_corrE} for the
topological oscillator. After exchanging the variables in
eq.~\eqref{equ:topo_corr} by the ones of the anharmonic oscillator
model a similar calculation has to be done here.

The harmonic and anharmonic quantum mechanical oscillators have been 
studied with QMC methods in \cite{Jansen:2012gj,Jansen:2013jpa,Ammon:2013yka}.

\section{Remarks on Quasi-Monte Carlo attempts}\label{sec:QMC}
In this section, we would like to give an overview of our attempts to
solve the topological oscillator with QMC methods utilizing
(randomized) Sobol' sequences (cf., \cite{SOB67, OWE95})
\footnote{These Sobol' sequences are low-discrepancy sequences, i.e.,
  they are designed to be more uniformly distributed than
  pseudo-random numbers.}. The section is intended for researchers
who are familiar with QMC and may be skipped by non-experts. The addressed attempts were motivated by the
positive results obtained for the (an)harmonic oscillator model (cf.,
e.g., \cite{Jansen:2012gj}). At the present state, however, we were
not able to observe positive results with our selected QMC
constructions for the topological oscillator model. In the following,
we will summarize the sampling techniques that were tested in
combination with (randomized) Sobol' sequences.

\paragraph{Na\"ive Sampling} The na\"ive sampling here simply means that we used the (randomized) Sobol' 
points as an approximate uniform distribution and used these points directly to evaluate the (re-scaled) integrals according to
\begin{align*}
  \int_{[0,1]^d}f(x)dx\approx\frac1N\sum_{i=1}^Nf(x_i)
\end{align*}
where the $x_i$ are the $d$-dimensional Sobol' points. As to be expected, it produced good results for small 
$\frac Ia$ but loses accuracy and convergence speed as $\frac Ia$ increases since most samples have little to 
no weight rendering them irrelevant. In fact, as $\frac Ia$ grows significantly larger than $\frac12$ we have not 
been able to observe any convergence in the range of sample sizes that are feasibly generatable.

\paragraph{Harmonic Sampling} Since we observed good results using the Sobol' sequences with the harmonic 
and anharmonic oscillator, it seemed reasonable to approximate 
\begin{align*}
  \cos\left(\phi_i-\phi_j\right)\approx 1-\left(\phi_i-\phi_j\right)^2,
\end{align*}
i.e., to draw just as in \cite{Jansen:2012gj} and re-weight to account for the error made in drawing from this 
approximate distribution. Using the cumulative (standard) normal distribution $\Phi$, there are three obvious 
approaches to inverting $\Phi$ with range $[-\pi,\pi)$.
\begin{itemize}
\item Using $\mathbb{R}$ as a covering space of $[-\pi,\pi)$ yields the ``inverse''
  \begin{align*}
    (0,1)\to[-\pi,\pi);\ x\mapsto\Phi^{-1}(x)\ \operatorname{MOD}\ 2\pi
  \end{align*}
  where
  \begin{align*}
    \operatorname{MOD}:\ \mathbb{R}\to[-\pi,\pi);\ x\mapsto x-\operatorname{round}\left(\frac{x}{2\pi}\right)2\pi=x-\left\lfloor 
    \frac{x}{2\pi}+\frac12\right\rfloor2\pi.
  \end{align*}
\item Use the complete normal distribution and dismiss points not in $[-\pi,\pi]^d$. 
(Note that this approach destroys the property of uniformity of the Sobol' sequence by rejecting some of the points.)
\item Restrict to $[-\pi,\pi)$-slice of the normal distribution, i.e., invert 
  \begin{align*}
    \tilde\Phi(x):=\frac{\Phi(x)-\Phi(-\pi)}{2\Phi(\pi)-1}.
  \end{align*}
\end{itemize}
Qualitatively, all three approaches yield the same result; our quadrature was highly 
instable yielding seemingly random numbers and we have not been able to stabilize them. 
We think this is due to an inherent under-representation of samples in the region where 
$\left\vert\phi_{i+1}-\phi_i\right\vert$ is close to $2\pi$. These samples have a large weight 
but the chance of drawing them is slim. Hence, hitting these regions a little more often can make a significant 
difference which we observe as an instability of our results.

\paragraph{Inversive Samplings} Inversive samplings (cf., \cite{Devroye} chapter II.2) go one step 
further than the harmonic sampling by choosing better approximations than the harmonic one. 
However, we still have the constraint that we need to be able to effectively draw from the distribution. 
In this case, we are looking for an expression of the form
\begin{align*}
  \exp\left(-S[\phi] \right)\approx\prod_{j=0}^{d-1}p_j\left(\phi_j\right).
\end{align*}
In fact, the two main examples we tested were of the form
\begin{align*}
  \exp\left(-S[\phi] \right)\approx\prod_{j=0}^{d-1}p\left(\phi_j\right)
\end{align*}
where $p$ is a polynomial or a step-function. The step-function has the advantage that we can draw with very little effort. 
However, QMC does not like discontinuous integrands, i.e., we expect to lose convergence speed. Choosing a polynomial 
$p$ does not have that draw-back but generating samples is a lot harder since it involves numerically 
inverting the cumulative distribution function defined by the polynomial used as a density.\\

Our results are satisfactory in the sense that we did not observe an increase in autocorrelation time going to small 
lattice spacings (in fact, we would be highly surprised if that happened since we are not using a random walk to generate samples) 
but we have not been able to improve over standard Monte Carlo methods where they are applicable, 
that is, to obtain an error scaling better than $O\left(N^{-0.5}\right)$ where $N$ is the sample size. 

\paragraph{Sampling - Importance Resampling} SIR (cf., \cite{Rubin87,Rubin88,GordonSalmondSmith93}) is a closely 
related concept where a pool of sample points is generated (here, using (randomized) Sobol' sequences) and stored. 
Samples are then to be drawn from the discrete distribution of points in the pool with respect to 
their weight in the target distribution (cf., e.g., \cite{Devroye} chapter III.2). 
It worked fairly well for small sample sizes but, trying to go to large numbers of samples, 
the pool must increase as well in order to keep the systematic error of the changed distribution 
small rendering this method too memory demanding to be viable for high accuracy calculations (an error analysis of QMC based SIR methods can be found in \cite{QMC-SIR}).

\paragraph{Envelope Inversive Rejective Samplings} An approach to draw directly from the target 
distribution but with the help of inversive sampling starts by choosing an approximation $\tilde S [\phi]$ of $S[\phi]$ such that 
\begin{align*}
  e^{-S[\phi]}\le e^{-\tilde S[\phi]}.
\end{align*}
We call such an inversive sampling an Envelope Inversive Sampling. Drawing from the envelope, 
we may now use a rejection step (cf., \cite{von-Neumann}) to ensure drawing from the target distribution 
in the usual Monte Carlo manner.\\

Our results are comparable to the Inversive sampling results, that is, no increased auto-correlation 
for small lattice spacings but also no accelerated convergence with respect to standard Monte-Carlo methods.

\paragraph{Smoothed Envelope Inversive Rejective Samplings} Here, the idea is that we might obtain an 
edge starting from the envelope inversive rejective sampling by smoothing the integrand; viz., the rejective step is a step-function in the 
integrand which behaves poorly with QMC and, hence, may yield better results if subjected to a smoothing operation. 
The method of Smoothed Rejection can be found in \cite{CafMos96} and effectively replaces the step-function 
describing the rejection step by a continuous function and additional re-weighting. However, we could not observe any 
improvement over the envelope inversive rejective sampling.

\section{The method of recursive numerical integration}\label{sec:Recursive}
In this section, we consider the method of \textit{recursive numerical integration} (see \cite{Genz86,Hayter06} and references therein) 
also sometimes called the method of \textit{iterated numerical integration}, 
for the approximation of the integration problems in 1-dimensional lattice theory stated in section \ref{sec:1D-Model}. 
The main idea of this method is based on the fact that if the
integrand at hand can be described as the product 
of low-dimensional functions, each one describing the interactions or \textit{couplings} between few 
consecutive objects, then we can write the final high-dimensional integration problem as the iteration of 
coupled low-dimensional integrals. At this point a quadrature rule for low-dimensional integration has to be chosen 
in order to carry out the successive approximation of the underlying low-dimensional integrals. In the following, we will consider 
the case of simple 1-neighboring couplings for simplicity. For a description of the general case with several neighboring couplings 
or branchings, we refer again to \cite{Hayter06}. Thus, we have an integrand function of the form
\begin{align*}
  f(x)=\prod_{i=0}^{d-1} f_{i}(x_i,x_{(i+1) (d)})
\end{align*}
and we would like to approximate 
\begin{align*}
  I= \int_{D^d}\prod_{i=0}^{d-1} f_{i}(x_i,x_{(i+1) (d)}) d\vx, 
\end{align*}
where the notation $(.)(d)$ means to take the argument modulo $d$, and $D\subset \mathbb{R}$ is a 1-dimensional domain. 
Thus, by the Fubini-Tonelli theorem, assuming that the desired integral value exists, we can write the problem equivalently as
\begin{footnotesize}
\begin{align*}
I=& \int_{D}\Big( \int_{D} f_{0}(x_{0},x_{1}) \dots   \\
&\dots \left(\int_{D}  f_{d-3}(x_{d-3},x_{d-2}) \left(\int_{D}   f_{d-2}(x_{d-2},x_{d-1}) f_{d-1}(x_{d-1},x_{0}) dx_{d-1} \right) dx_{d-2} \right)\dots \\
&\dots dx_1\Big)dx_{0}. 
\end{align*} 
\end{footnotesize}
Furthermore, we can consider scalings $c_0, \dots, c_{d-1} >0$, and define $I^{\star}:=\left(\prod_{i=0}^{n-1}\frac{1}{c_i}\right) I$,  such that
\begin{footnotesize}
\begin{align*}
I^{\star}=& \int_{D}\Big( \int_{D} \frac{f_{0}(x_{0},x_{1})}{c_0} \dots  \\ 
&\dots  \left(\int_{D}  \frac{f_{d-3}(x_{d-3},x_{d-2})}{c_{d-3}} 
  \left(\int_{D}  \frac{ f_{d-2}(x_{d-2},x_{d-1}) f_{d-1}(x_{d-1},x_{0})}{c_{d-2} c_{d-1}} dx_{d-1} \right) dx_{d-2} \right) \dots\\
  &\dots dx_1\Big)dx_{0}. 
\end{align*}
\end{footnotesize}
Usually the quantities $c_0, \dots, c_{d-1}$ will have to be chosen adaptively as they are used in order to avoid under/over-flows in the recursive 
method for high dimensions due to limited machine accuracy. By selecting an adequate 1-dimensional quadrature rule $Q$ with 
$m$ points and weights for approximating the underlying integration problems over $D$ in each iteration, we can write the iteration method in 
recipe form as
\begin{enumerate}
 \item Fix, if possible, a quadrature rule with $m$ points $x^1,\dots,x^m$ and weights $w_1,\dots,w_m$ that works well for one dimensional integrals of type
   \begin{align}
     \int_{D} f_{i-1}(\theta_1, z)f_i(z,\theta_2) dz, \quad \text{ for
     } 1\le i \le d-1, \:  \theta_1,\theta_2 \in D.    
     \label{param_inte}
   \end{align}
\item Use the quadrature in 1. to estimate
\begin{align*}
 F_{d-1}(x_{d-2},x_{0})&:=\frac{1}{c_{d-2}c_{d-1}}\int_{D}   f_{d-2}(x_{d-2},x_{d-1})f_{d-1}(x_{d-1},x_{0}) dx_{d-1} \\
 &\approx \frac{1}{c_{d-2}c_{d-1}} \sum_{j=1}^{m} w_j f_{d-2}(x_{d-2},x^j)f_{d-1}(x^j,x_{0}),
\end{align*}
 over the grid of points $\{x^1,\dots,x^m\}\times\{x^1,\dots,x^m\}\subset D^2$. By defining the \textit{transfer} 
$m\times m$ matrix 
 \begin{align*}
   M_{i}(k,l):=f_{i}(x^k,x^l), \quad 1\le k,l \le m, \quad 0\le i \le d-1,
 \end{align*}
we can write in matrix form 
\begin{align*}
  [F_{d-1}(x^i_{d-2},x^j_{0})]_{1\le i,j \le m} \approx \frac{1}{c_{d-2}c_{d-1}}  M_{d-2} diag((w_1,\dots,w_m)) M_{d-1},
\end{align*}
where 
\begin{align*}
  diag((w_1,\dots,w_m)):=
  \begin{pmatrix}
    w_1 & 0 & \cdots & 0 \\
    0 & w_2 & \cdots & 0 \\
    \vdots  & \vdots  & \ddots & \vdots  \\
    0 & 0 & \cdots & w_m
  \end{pmatrix}.
\end{align*}
\item For $i=d-2,d-3, \dots,2$ approximate iteratively
\begin{align*}
 F_{i}(x_{i-1},x_0)&= \frac{1}{c_{i-1}}\int_{D} f_{i-1}(x_{i-1},x_{i}) F_{i+1}(x_{i},x_0) dx_{i}\\
 &\approx \frac{1}{c_{i-1}} \sum_{j=1}^{m} w_j f_{i-1}(x_{i-1},x^j)F_{i+1}(x^j,x_0)
 \end{align*}
 over the grid of points $\{x^1,\dots,x^m\}\times\{x^1,\dots,x^m\}\subset D^2$. 
Thus, the result of step 3. over $\{x^1,\dots,x^m\}\times\{x^1,\dots,x^m\}$ can be written as the matrix products
\begin{align*}
  [F_{2}(x^i_1,x^j_0)]_{1\le i,j \le m}\approx \Big(\prod_{i=1}^{d-2}\frac{1}{c_{i}} M_{i} diag((w_1,\dots,w_m)) \Big) \frac{1}{c_{d-1}} M_{d-1},  
\end{align*}
with the notation for matrix products $\prod_{i=1}^{d-2}B_i:=B_1B_2\dots B_{d-2}$.
\item Estimate with the $m$-point quadrature in 1. the function
  \begin{align*}
    F_{1}(x_{0},x_0)=\int_{D} \frac{f_{0}(x_{1},x_{0}) F_{2}(x_{1},x_0)}{c_{0}}  dx_{1}.
  \end{align*}
over the grid $\{x_{0}^1,\dots,x_{0}^m\} \subset D$. 
Finally, estimate with the $m$-point quadrature in 1. the function
\begin{align*}
  I^{\star}=\int_{D} F_{1}(x_{0},x_0) dx_{0}.
\end{align*}
The result of this step can be written in matrix operations as 
\begin{align*}
 I^{\star}& \approx \mathrm{Tr} \left( diag((w_1,\dots,w_m))  
 \Big(\prod_{i=0}^{d-2}\frac{1}{c_{i}} M_{i} diag((w_1,\dots,w_m)) \Big) \frac{1}{c_{d-1}} M_{d-1} \right)\\
 &=\mathrm{Tr} \left( \prod_{i=0}^{d-1}\frac{1}{c_{i}} M_{i} diag((w_1,\dots,w_m)) \right).
\end{align*}
\end{enumerate}

As mentioned in the previous section we are mainly interested in considering two types of integrands for high-dimensional integration. 
The first type of integrand is given as the weight of the Boltzmann distribution. In this case we usually obtain that the functions $f_i, 0\le i\le d-1, $ satisfy $f_0=f_1=\dots=f_{d-1}$ due to isotropic conditions of the model. 
Thus, this case yields $M_0=M_1=\dots=M_{d-1}=:M$ and we can choose 
$c_0=c_1=\dots=c_{d-1}=:c$. Hence, the identity above reduces to

\begin{align*}
  I^{\star}=\mathrm{Tr} \left( \Big(\frac{1}{c} M diag((w_1,\dots,w_m)) \Big)^d \right).
\end{align*}
Note that the similarity relation

\begin{align*}
  M diag((w_1,\dots,w_m))\sim diag((\sqrt{w_1},\dots,\sqrt{w_m})) M diag((\sqrt{w_1},\dots,\sqrt{w_m})), 
\end{align*}
holds, where for $m\times m$ matrices $A,B$  we say that $A\sim B$ if and only if $A=CBC^{-1}$, for an invertible matrix $C$. 
The resulting matrix on the right hand side of the relation 
above is symmetric and efficient algorithms are known for its diagonalization (cf., \cite{Saad11}). 
By use of the eigenvalue decomposition of a diagonalizable matrix $A$, we can write $A^d=C D^d C^{-1}$ where $D$ is the diagonal matrix  
of eigenvalues, and $C$ is the corresponding invertible matrix of eigenvectors. 
Because the trace of a diagonalizable matrix equals the sum of its eigenvalues, 
to calculate the trace of $A^d$ we only need to calculate the 
eigenvalues of $A$, raise them to the power of $d$, and finally sum them up.
It is also worth to mention that for particular applications, \textit{recursive-multiplication} for 
calculation of $\Big(\frac{1}{c} M diag((w_1,\dots,w_m)) \Big)^d$ may be also a competitive method.\\
The second type of integrand is given by the Boltzmann weight times an observable function. For simple observable functions usually 
considered in lattice theory, similar gains over direct matrix-matrix multiplications can be obtained by condensing a sequence of 
matrix-matrix multiplications into a power form. Particular examples and the corresponding implementations will be discussed in detail in the 
following section.

\section{Numerical experiments}\label{sec:Num_Exp}
In this section, we will investigate the applicability of the method
of recursive numerical integration to the topological rotor 
and the anharmonic oscillator, as described in
section \ref{sec:1D-Model}. In particular, 
we will compare the calculations of recursive numerical integration
methods with standard MCMC methods applied
to lattice systems \cite{Gattringer:2010zz,DeGrand:2006zz} and also 
with improved methods, i.e., the cluster algorithm 
\cite{Niedermayer:1996ea} and randomized QMC methods for the 
anharmonic oscillator \cite{Jansen:2013jpa}.

\subsection{Topological oscillator}
For the sake of illustrating the difficulties that can appear in 
standard MCMC calculations,
we will perform initial computations with the Metropolis algorithm
\cite{Metropolis}. This algorithm, although very generally 
applicable, is often not the optimal choice to simulate a model and 
is usually replaced by a more suitable technique. 
However, here we use the Metropolis algorithm 
to demonstrate the generic difficulty of many  
MCMC methods that towards the continuum limit 
the autocorrelation time grows rapidly. 
At fixed number of samples, 
this leads to very large errors when the lattice spacing is 
reduced and, correspondingly, to highly increasing computational costs of
the simulations. Throughout the discussion 
below, we use a moment of inertia
of $I = 0.25$ for the quantum mechanical topological rotor.
 
Besides promising a much improved error scaling, 
recursive numerical integration methods avoid 
this explosion of the autocorrelation time 
and are therefore highly superior to MCMC techniques.            
Nevertheless, for special situations there exist  
MCMC methods which also avoid the increase in autocorrelation 
time. For the topological rotor investigated here, 
cluster algorithms \cite{Niedermayer:1996ea} can be applied which 
exhibit a basically constant behavior of the autocorrelation time
with respect to the lattice spacing while still showing the $1/\sqrt{N}$ 
error scaling behavior of MCMC methods. 
Hence, we will explore how the 
recursive numerical integration technique 
compares to an optimal MCMC method such as the cluster algorithm
and whether a gain can still be found. 
Since for the quantum mechanical topological rotor we
could not find a 
satisfactory realization
of QMC methods to solve the system, we will not discuss this 
approach in this section.


\subsubsection{MCMC method}

A basic algorithm of the MCMC class is the Metropolis algorithm
\cite{Metropolis} which, as discussed above, will be used 
for illustration purposes only. 
The Metropolis algorithm uses
importance sampling as described in section \ref{sec:QMC} to create
samples of a given system on which observables $O$ 
can be calculated. These observables are then 
averaged over all samples to give the expectation value 
$\langle O\rangle$ with an error $\Delta\langle O\rangle\propto 1/\sqrt{N}$, where $N$ 
is the number of samples. The problem of the Metropolis algorithm 
- and MCMC methods in general - is
that the samples are not all independent of each other but 
correlated. This correlation is measured through the 
{\em autocorrelation time} which is a property of the employed
algorithm and, if the so-called integrated autocorrelation 
time is taken, also of the considered observable. 

\begin{figure}[t]
  \includegraphics[width=0.49\textwidth, page=4]
  {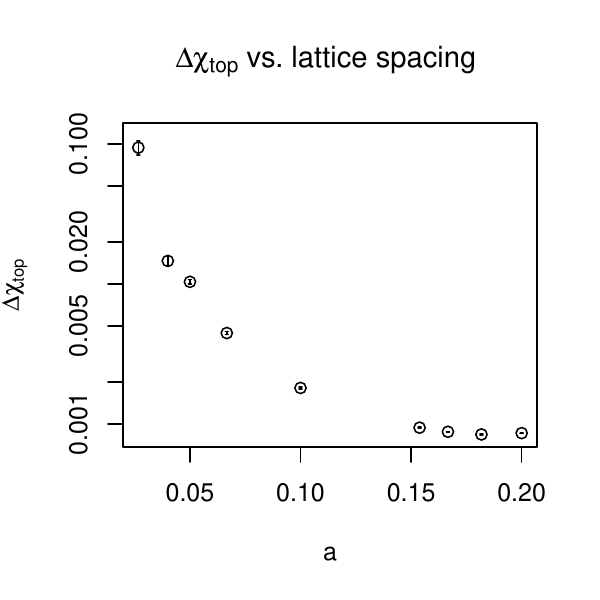} 
  \includegraphics[width=0.49\textwidth, page=5]
  {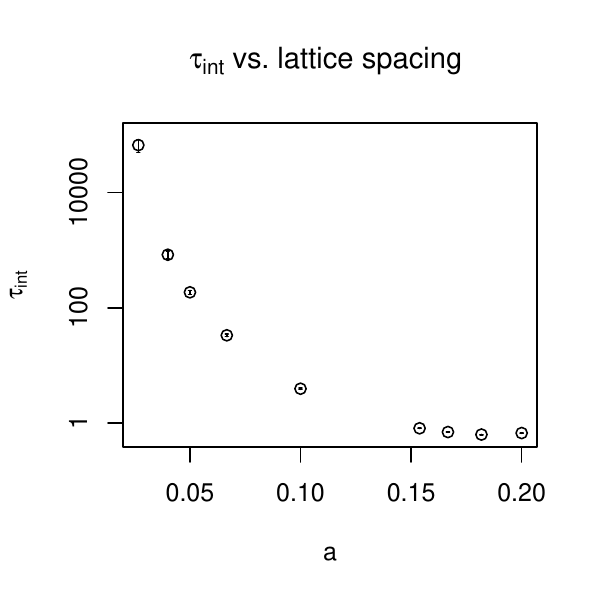}
  \caption{Logarithmic error and logarithmic integrated autocorrelation time behavior 
    of the Metropolis algorithm for fixed number of samples
    ($N = 10^5$) dependent on the lattice spacing $a$. Left: Error of
    the topological susceptibility $\Delta
    \chi_{\text{top}}$, right: the corresponding integrated autocorrelation time
    $\tau_{int}$.}
  \label{fig:metro-err}
\end{figure}

The behavior of the error 
and the integrated autocorrelation time of the topological susceptibility 
when applying the  
Metropolis algorithm as a function of the 
lattice spacing $a$ 
is shown in figure~\ref{fig:metro-err}. 
We have chosen the topological susceptibility as an observable, 
since it is very sensitive to 
correlations between samples and often shows the largest 
integrated autocorrelation time in a simulation.
It is also for this reason that the quantum mechanical topological 
rotor, where topological effects can be studied, is a good test case 
to compare MCMC and QMC or recursive numerical integration 
methods.

We remark that for all values 
of the lattice spacing shown in fig.~\ref{fig:metro-err} a fixed number of samples 
($N = 10^5$)
has been used. 
In addition, for the error calculation the 
integrated 
autocorrelation time has been fully taken into account. 
As can be observed in fig.~\ref{fig:metro-err}, 
the integrated autocorrelation time $\tau_{int}$ (right plot)
and therefore the error of the topological susceptibility $\Delta
\chi_{\text{top}}$ (left plot) are rapidly growing with decreasing lattice spacing
$a$.
This leads to highly increased computational costs. 

For  
systems in statistical physics or quantum field theory the 
continuum limit is reached by approaching a critical point. 
Since the
rapid increase of the autocorrelation time in this limit    
is generic for many MCMC algorithms, this  
constitutes a most severe problem when higher dimensional 
systems are explored.

\begin{figure}[t]
  \includegraphics[width=0.49\textwidth, page=2]
  {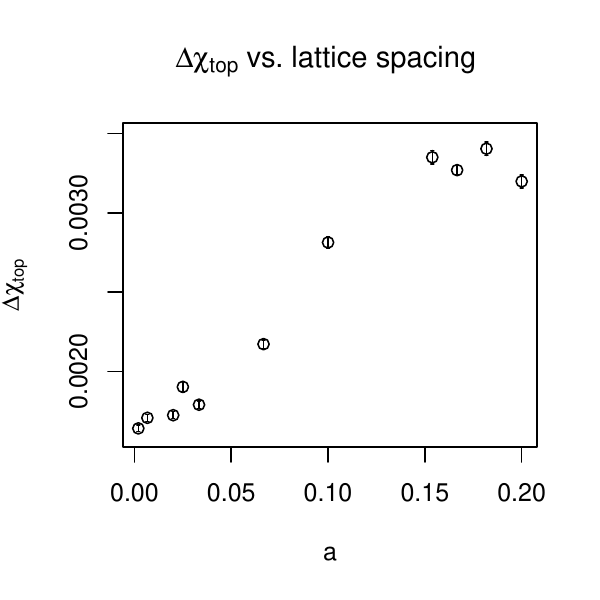} 
  \includegraphics[width=0.49\textwidth, page=3]
  {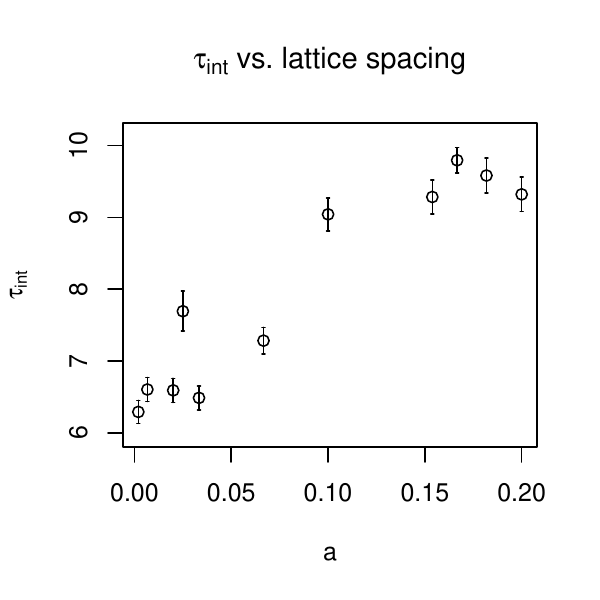}
  \caption{Error and integrated autocorrelation time behavior
    of the cluster algorithm for fixed number of samples 
    ($N = 10^5$) dependent on the lattice spacing $a$. Left: Error of
    the topological susceptibility $\Delta
    \chi_{\text{top}}$, right: the corresponding integrated autocorrelation time
    $\tau_{int}$.}
  \label{fig:cluster-err}
\end{figure}

For the explicit model considered here, there exists a 
cluster algorithm \cite{Niedermayer:1996ea} which avoids the 
problem of an increasing autocorrelation time for 
shrinking lattice spacings. 
Figure \ref{fig:cluster-err}
shows the behavior of the autocorrelation
time (right plot) and the error of the topological susceptibility
(left plot) for the cluster algorithm as a function of $a$. 
As can be seen for the cluster algorithm, there is even a tendency
that both, the error and the autocorrelation time, shrink for smaller
$a$.

\begin{figure}[h!]
  \centering
  \includegraphics[width=.6\textwidth]
  {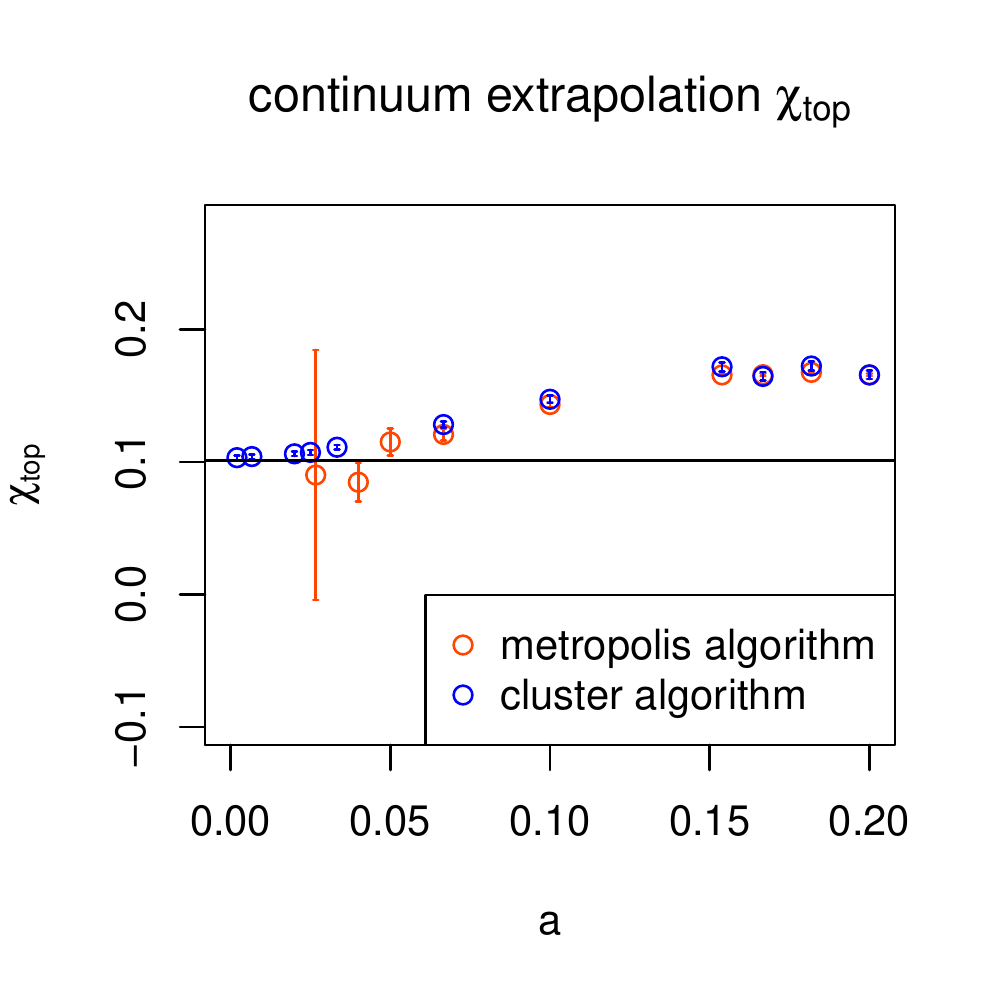} 
  \caption{Continuum extrapolation of the topological susceptibility
    $\chi_{\text{top}}$ calculated by the Metropolis and the cluster algorithms
    for a fixed number of samples ($N = 10^5$).}
  \label{fig:mcmc-chi}
\end{figure}

To compare the two algorithms directly, figure \ref{fig:mcmc-chi}
shows the observable $\chi_{\text{top}}$ dependent on $a$ for 
the Metropolis and the cluster algorithms.  
For larger values of $a$ the error from both algorithms are similar. 
However, for
small values of $a$ the error of the Metropolis algorithm becomes so large that
it would be very demanding 
to reach an accurate result 
below a certain value of the lattice spacing, thus
making a continuum extrapolation of $\chi_{\text{top}}$ rather difficult. 
On the other hand, for the cluster algorithm the error stays practically 
constant; hence, allowing us to reach very small values of the lattice spacing 
giving a much better control of the continuum limit. 

We stress again that the discussion above is only intended for an illustration 
of the generic behavior of MCMC methods and to demonstrate the problem 
of an increasing autocorrelation time when approaching the continuum
limit for certain classes of MCMC algorithms. As the examples show, MCMC methods
often have to face the difficulty of very large autocorrelation times which 
would be absent for QMC or recursive numerical integration methods. 
In the model considered here an optimal algorithm can be used, the cluster
algorithm. It will therefore be interesting to see, whether the 
recursive numerical integration method is still advantageous even for cases
when highly improved MCMC techniques can be employed. 
Although not relevant for this paper,
we would like to mention that 
cluster algorithms are not applicable to gauge theories, so far.

\subsubsection{Recursive Gaussian quadrature}

\begin{figure}[h!]
  \centering
  \setlength{\columnsep}{-3cm}
  \begin{multicols}{2}
    \includegraphics[width=.4\textwidth, page=3]
    {plots/results_all_m120.pdf}
    \includegraphics[width=.4\textwidth, page=4]
    {plots/results_all_m120.pdf}
  \end{multicols}
  \vspace{-1cm}
  \begin{multicols}{2}
    \includegraphics[width=.4\textwidth, page=5]
    {plots/results_all_m120.pdf}
    \includegraphics[width=.4\textwidth, page=6]
    {plots/results_all_m120.pdf}
  \end{multicols}
  \vspace{-1cm}
  \begin{multicols}{2}
    \includegraphics[width=.4\textwidth, page=1]
    {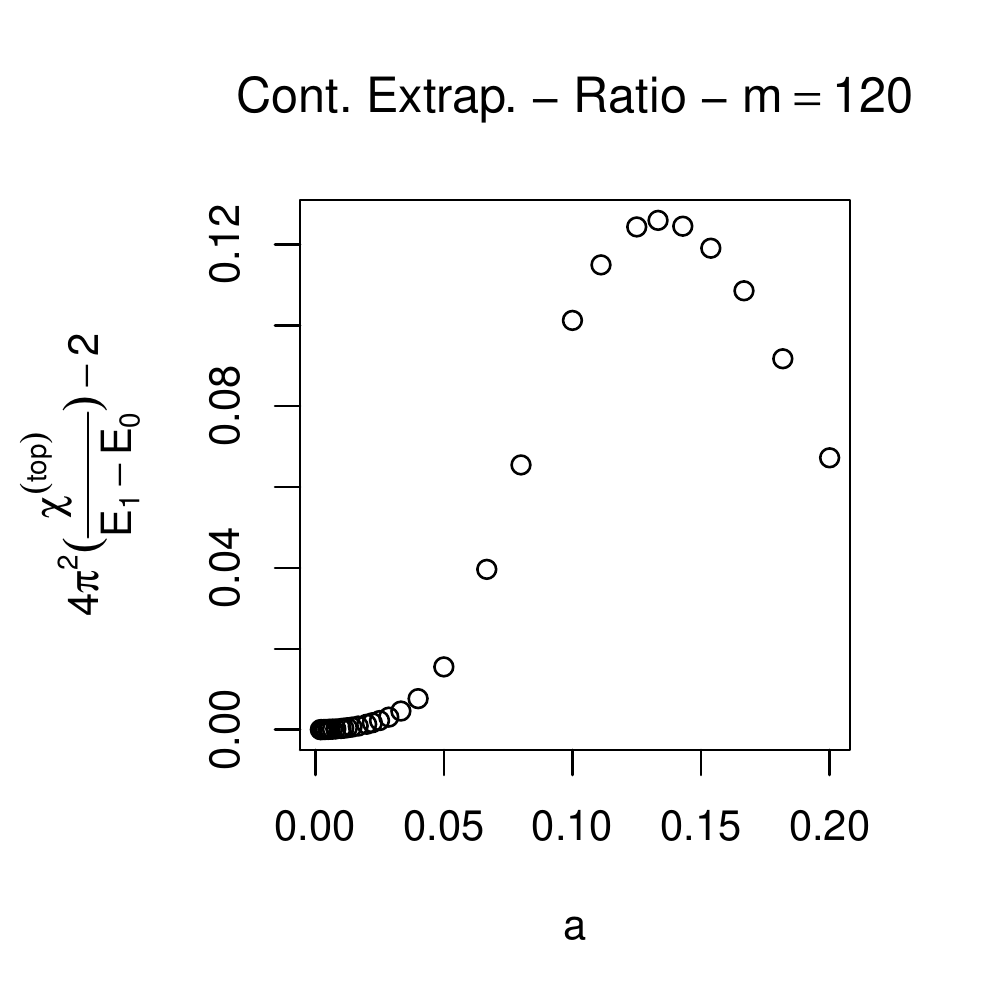}
    \includegraphics[width=.4\textwidth, page=2]
    {plots/results_all_m120.pdf}
  \end{multicols}
  \vspace{-.5cm}
  \caption{Gaussian quadrature - continuum extrapolation of three
    different observables, the topological susceptibility
    $\chi_{\text{top}}$ (top row), the energy gap $\Delta E = E_1 -
    E_0$ (middle row) and the ratio $\frac{\chi_{\text{top}}}{\Delta
      E}$ (bottom row) with $m=120$ integration points. Each
    observable is presented with a linear axis scale (left column) and a
    logarithmic axis scale (right column). For all results shown, we have
    subtracted the theoretical value of the given observable in the
    continuum such that they should converge to zero in the continuum
    limit.}
\label{fig:gaussobservables}
\end{figure}

In order to demonstrate the precision that can be reached with 
the recursive numerical integration method even at very small 
values of the lattice spacing, we will first discuss 
some numerical results that we have obtained with this approach. 
To this end, 
we have implemented the method of recursive numerical integration described in section
\ref{sec:Recursive}, using
Gauss-Legendre mesh points (abscissae), and applied it to the topological
oscillator.
Fixing the number of integration points to $m=120$ and the value of the moment of inertia
$I=0.25$, we calculated 
the topological susceptibility  
$\chi_{\text{top}}$, the energy gap $\Delta E$ 
and the ratio of both observables
$\frac{\chi_{\text{top}}}{\Delta E}$ as a function 
of the lattice spacing; cf., fig.~\ref{fig:gaussobservables}. 

For the employed value of $I=0.25$, there are 
theoretical predictions for the observables we consider here, 
see section \ref{sec:1D-Model}. In particular, in the 
continuum limit we should find for
the topological susceptibility
$\chi_{\text{top}} \rightarrow \frac{1}{4\pi^2I} = \frac{1}{\pi^2}$, for the 
energy gap 
$\Delta E \rightarrow \frac{1}{2I} = 2$, and for the 
ratio 
$\frac{\chi_{\text{top}}}{\Delta E} \rightarrow \frac{1}{2 \pi^2}$. 

In figure~\ref{fig:gaussobservables}, we show our results obtained 
with recursive Gauss-Legendre quadrature with the expected 
continuum values, as given above, subtracted. 
The graphs nicely show that for all observables the results 
converge to zero; thus, being fully consistent with the 
theoretical expectations.  
Note that in the left column of fig.~\ref{fig:gaussobservables} 
we use a linear scale for the observables considered 
while in the right column a logarithmic scale is used which 
demonstrates the high precision we can reach with recursive Gauss-Legendre quadrature. 
Furthermore, note that the ratio $\frac{\chi_{\text{top}}}{\Delta E}$ has a peak at higher values of $a$, due to peculiar 
lattice artifacts, which is consistent with previous
studies of this model \cite{Bietenholz:2010xg}.

\begin{figure}[h]
  \centering
  \includegraphics[width=.6\textwidth]
  {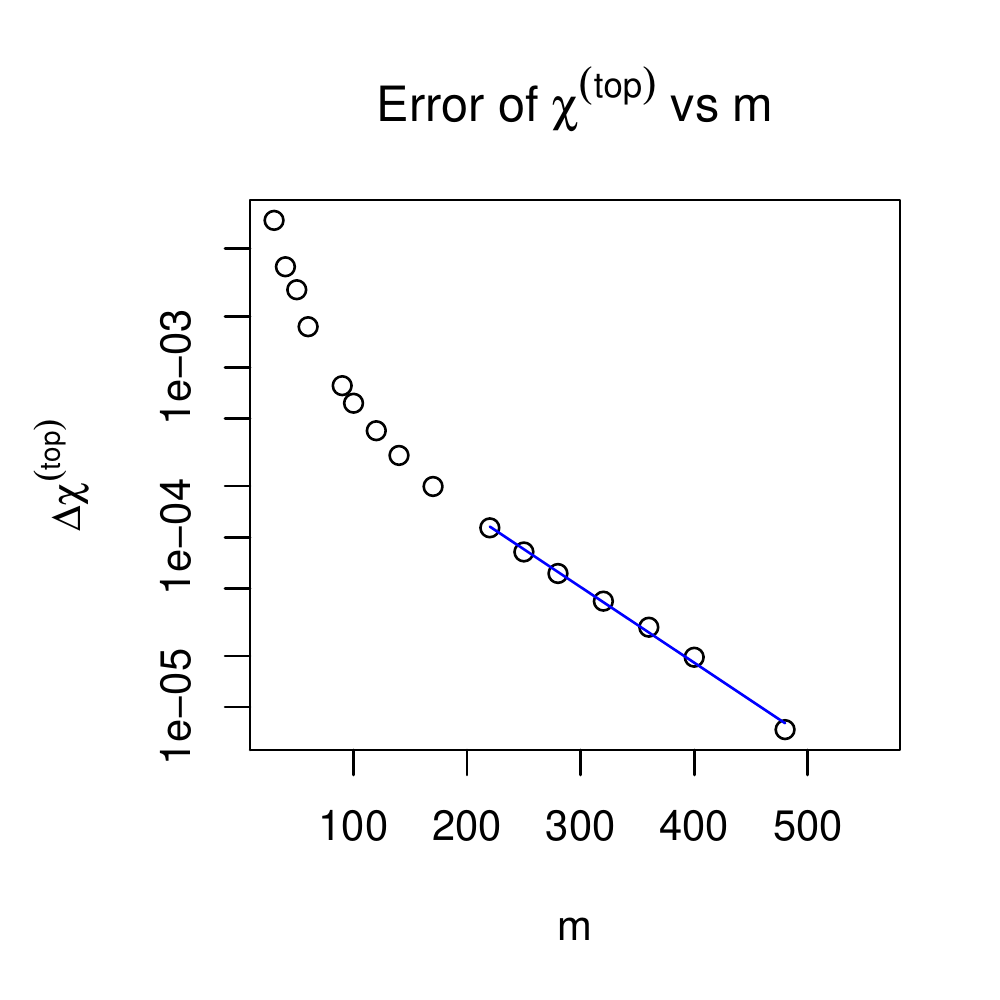} 
  \caption{Gaussian quadrature - error behavior of
    $\chi_{\text{top}}$ in dependence of the number of points $m$ used
    in the integration and with fixed lattice constant $a = 0.4$. Note 
    that we use a logarithmic scale to exhibit the error. The
    blue line is an exponential fit of the error for $220 \le m \le 480$ which 
    appears - through the use of a logarithmic scale - 
    as a straight line  
    for a function of $m$. 
    The consistency
    of the data with the linear behavior suggest that the error 
    scales down at least exponentially fast.}
  \label{fig:gauss-dchi-m}
\end{figure}

Let us now turn to the interesting question of the error scaling
for the recursive integration method. 
The behavior of the error for the topological 
susceptibility is shown in figure \ref{fig:gauss-dchi-m}
for a fixed lattice spacing $a=0.4$ as a function of the number
of integration points $m$.
Note that we use a logarithmic scale for plotting the 
error of the topological susceptibility. 
We define the error by the difference of $\chi_{\text{top}}$ 
obtained for $m=560$ and $\chi_{\text{top}}$ computed at the 
given number of integration points $m \le 480$. 
For $220 \le m \le 480$ we fit an exponential 
function of the error in $m$ which appears as a straight line 
in fig.~\ref{fig:gauss-dchi-m} where a logarithmic 
scale is used. The good agreement between the data and this 
exponential fit suggests that asymptotically the error 
scales down at least exponentially fast. Recall that the error behavior
for Gauss-Legendre quadrature with $m$ points for infinitely differentiable functions  
scales like $O(\frac{1}{(2m)!})\sim O(\frac{e^{2m}}{\sqrt{2\pi 2m} (2m)^{2m}})$, where the latter 
relation holds due to Stirling's approximation formula.

It is clear that with asymptotic exponential (or even better) error 
scaling the numerical recursive integration method 
will outperform any algorithm that shows an algebraic error scaling, 
in particular the $1/\sqrt{N}$ behavior of MCMC algorithms. 
Still, it is an interesting question whether for small, 
more practical values 
of $N$ (or $m$) a gain can be obtained from 
the recursive integration method. 

We, therefore, 
measured the run-time of the Gauss and the cluster algorithms
needed to obtain a given error of the topological susceptibility. 
For these measurements we ran both algorithms
with the same input parameters (we used here a lattice
  spacing of $a=0.1$) on the same stand-alone computer. We then varied
the number of samples and integration points for the cluster and recursive Gauss-Legendre quadrature,
respectively, and measured the run-time. We repeated each measurement ten
times and averaged over the measured times to get an error estimate of
the run-time. This error originates from the number and kind of other processes
that are running on the machine at a given time and is noticeable for small run-times. 
For the cluster algorithm, in addition, the size and distributions of the 
generated clusters can vary leading to different run-times of the algorithm. 

\begin{figure}[h]
  \centering
  \includegraphics[width=\textwidth]
  {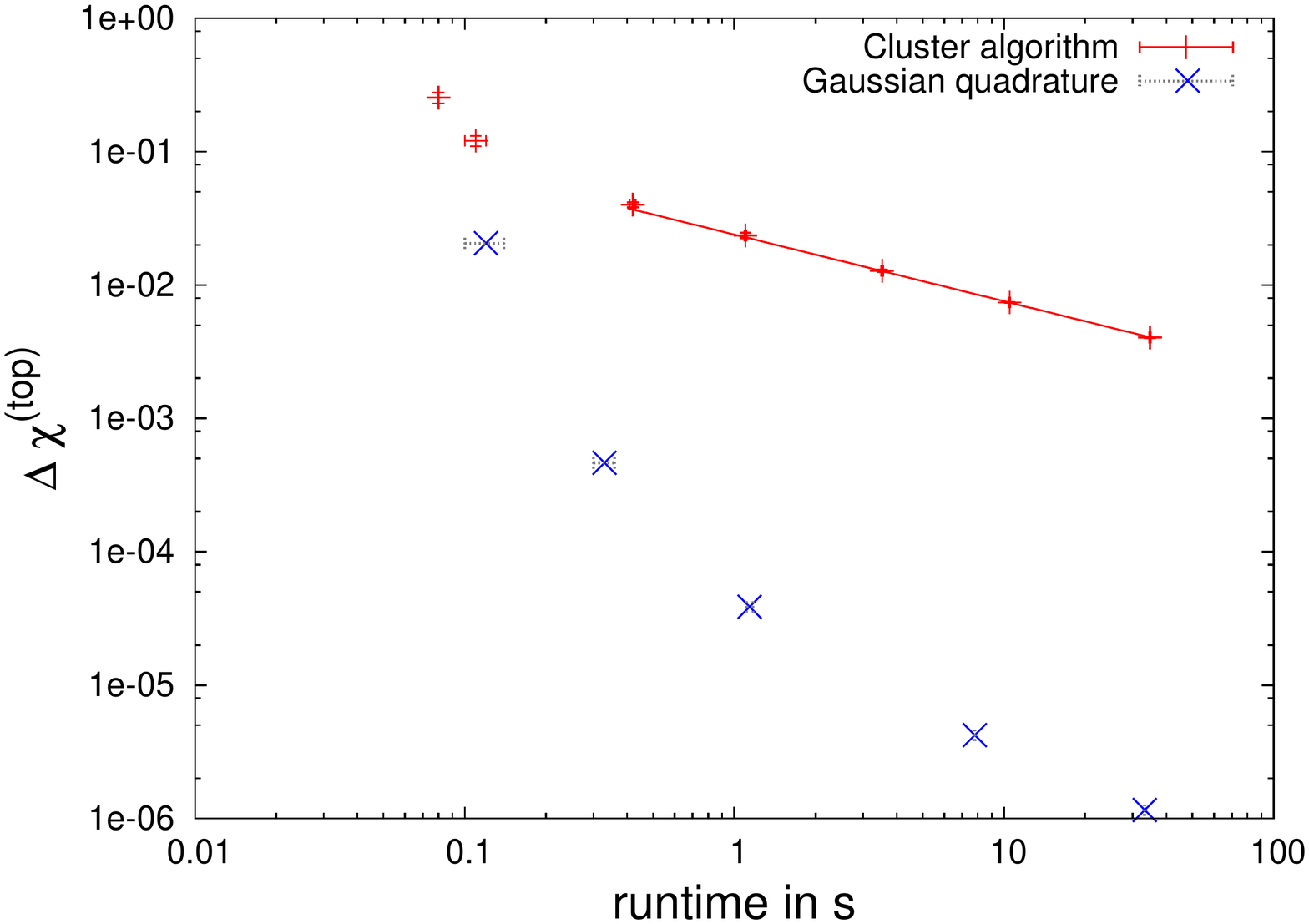} 
  \caption{Comparison of Cluster algorithm and recursive Gauss-Legendre quadrature. 
    We show the run-time $t$ in
    seconds needed to get a given error on the topological
    susceptibility with fixed lattice spacing $a = 0.1$ on a 
    stand-alone computer. Note that the graph is plotted in a double
    logarithmic scale. We used 
    $N = 10^2 ... 10^6$ samples for the cluster algorithm 
    and $m = 10 ... 300$ integration points for the Gauss
    algorithm. To achieve the error of the error 
    the numerical experiment has been repeated ten times for a fixed 
    set of parameters. The run-time error of both algorithms results 
    from different 
    execution times of the program due to state of the computer when 
    the program was executed. For the cluster algorithm the error 
    in the run-time is also determined from different distributions 
    of clusters generated.
    The red line shows the expected $1 / \sqrt{t}$ behavior
    of the cluster algorithm as a MCMC method.}
  \label{fig:dchi-t}
\end{figure}

Being an MCMC method, for the cluster algorithm the procedure of
repeating all runs ten times allows us to also determine the error on
$\Delta \chi_{\text{top}}$, i.e., the error of the error. The
recursive Gauss-Legendre quadrature, on the other hand, is purely
deterministic and, therefore, gives the same result for
$\chi_{\text{top}}$ without any error every time we run it. Since for
this test we have used different parameters, we have also chosen a
different gauge value of $m=400$ and calculate the error for
$\chi_i(m<400)$ via the difference to this value, $\Delta \chi_i =
|\chi_i - \chi_i(m=400)|$. The error of this error is neglected here.

In figure \ref{fig:dchi-t}, the run-time $t$ needed to achieve a given error of
the topological susceptibility is shown for both algorithms. 
Note that the graph is shown in a double logarithmic scale.
The cluster algorithm shows the for
MCMC methods typical $1 / \sqrt{t}$ behavior, indicated by the red
line. The recursive Gauss-Legendre quadrature has a much steeper negative
slope, such that the calculation to reduce the error by a specific
amount is substantially faster than using the cluster algorithm.
Additionally, even for small run-times the recursive Gauss-Legendre quadrature shows
a much reduced error already. Therefore, the recursive Gauss-Legendre quadrature 
is not only superior in the asymptotic regime at large run-times where it
shows a much improved error scaling but is also advantageous when 
only small run-times are employed. This makes it a promising approach if 
one thinks of 
systems in higher dimensions where only short run-times can be afforded. 

\subsection{Anharmonic oscillator}\label{ssec:ExperimentsAnharmonicOszi}
Applications of QMC methods to the anharmonic oscillator model and comparison with MCMC techniques have been studied in 
\cite{Jansen:2012gj,Jansen:2013jpa,Ammon:2013yka}. In particular, the superiority of randomized QMC techniques 
applied to the anharmonic oscillator model has been observed for small time periods $T\le 1.5$ independently of the 
size of lattice spacing $a$. Here we investigate whether the recursive integration 
method can also be successfully applied to this model and whether improvements over randomized QMC can be observed.
To this end, we carry out tests by choosing different pairs of problem dimension and spacing $(d,a)$.   
The time period $T$ associated to a pair $(d,a)$ is given by $T=da$. 
The choice of the pairs $(d,a)$ has been done with the aim to show that the 
anharmonic oscillator model can be approximated satisfactorily with a particular recursive numerical integration method. 
The observables investigated are $\langle x^2 \rangle$ and $\langle x^4 \rangle$. 
The rest of the parameters of the model have been fixed for the experiments to 
$M_0=0.5$, $\mu^2=-16.0$, and $\lambda=1.0$. We perform recursive one-dimensional integration 
based on Gauss-Legendre quadrature. To this end, we first re-scale the integration variables and re-arrange 
the terms in the action such that (using the same notation as in section \ref{sec:Recursive}) we end up with a transition function
\[
 f_i(x_i,x_{i+1})=e^{-(x_{i}-x_{i+1})^2  -
 \frac{1}{2} \left(\frac{\mu^2 a^2}{M_0}(x_{i}^2+x_{i+1}^2) + \frac{4 \lambda a^3}{M_0^2}(x_{i}^4 +x_{i+1}^4)\right)}.
\]
The quartic negative term $-\frac{4 \lambda a^3}{M_0^2}x_i^4$ dominates the quadratic positive term 
$-\frac{\mu^2 a^2}{M_0}x_i^2$ when $|x_i| \ge \sqrt{\frac{M_0 |\mu^2|}{4  \lambda a}}$, for $i=1,\dots,d$.
Outside of the region $\big[-\sqrt{\frac{M_0 |\mu^2|}{4  \lambda a}},\sqrt{\frac{M_0 |\mu^2|}{4  \lambda a}}\big]^2$, the 
term inside the exponential in the transition function $f_i(x_i,x_{i+1})$ remains always negative, and decays mainly 
with a quartic rate that adds to the quadratic coupling $-(x_{i}-x_{i+1})^2$. Therefore the marginal tails 
of this two-dimensional transition function decay faster than the marginal tails of a bivariate normal, and this fact can be used 
to select a region of main importance for the one-dimensional parametric integration problems described in \eqref{param_inte}. 
Thus, we additionally search to ensure that the decay obtained by the dominant quartic term overshadows the contribution 
of the remaining coupling quadratic term  $-(x_{i}-x_{i+1})^2$ in the action. 
As an heuristic, we then chose the main importance region for integration to be 
of the form $\big[-p\sqrt{\frac{M_0 |\mu^2|}{4  \lambda a}},p\sqrt{\frac{M_0 |\mu^2|}{4  \lambda a}}\big]$, 
for a number $p\ge1$, and validate numerically that the remaining integration region 
$\big(-\infty,-p\sqrt{\frac{M_0 |\mu^2|}{4  \lambda a}}\big] \cup \big[p\sqrt{\frac{M_0 |\mu^2|}{4  \lambda a}},+\infty\big)$ has a negligible 
contribution to the integration problem. Note that this is 
the same as to say that the original integration problem over $\mathbb{R}^d$ can be truncated satisfactorily for 
recursive numerical integration to the integration region 
$\big[-p\sqrt{\frac{M_0 |\mu^2|}{4  \lambda a}},p\sqrt{\frac{M_0 |\mu^2|}{4  \lambda a}}\big]^d$.
The selection of $p\ge 1$ can be taken to be a positive value that ensures  
$-\frac{\mu^2 a^2}{M_0}x_{i}^2 - \frac{4 \lambda a^3}{M_0^2}x_{i}^4 \le -K$, for some 
positive value $K$, and $x_i$ in $\big(-\infty,-p\sqrt{\frac{M_0 |\mu^2|}{4  \lambda a}}\big] \cup \big[p\sqrt{\frac{M_0 |\mu^2|}{4  \lambda a}},+\infty\big)$.
By integrating the tails of a normal density $\frac{1}{\sigma \sqrt{2 \pi}}e^{\frac{-t^2}{2 \sigma^2}}$ outside of the region $|t|\le 7\sigma$, 
we obtain a value below (but close to) $10^{-10}$. 
The maximal value of the exponential function in the density is given by $e^{-\frac{49}{2}}$ at $t=7\sigma$. Thus, for our tests we may select the conservative 
value of $K=24.5$ based on the fact that the tails of the one-dimensional integration problems in the 
recursive numerical integration method decay faster than 
a normal density, and that inside of the main importance region the behavior of the marginals seems mainly similar to the one of a 
shifted normal density with $\sigma=\frac{1}{\sqrt{2}}$ (due to the coupling quadratic term). 
This choice also makes sense since our computations are carried out in double precision and we aim to obtain 
at most 10 digits accuracy for the integration problems.
Thus, at the end, we select $p\ge 1$ to be the positive number that satisfies
\begin{align*}
-\left(\frac{\mu^2 a^2}{M_0}\left(p\sqrt{\frac{M_0 |\mu^2|}{4  \lambda a}}\right)^2 + 
\frac{4 \lambda a^3}{M_0^2}\left(p\sqrt{\frac{M_0 |\mu^2|}{4  \lambda a}}\right)^4\right) 
&=\left(p^2-p^4\right)\left(\frac{|\mu^2|^2 a}{4 \lambda} \right)\\
&= -24.5.
\end{align*}
We choose to take sample sizes $m$ inside of the selected importance region to be 
small multiples of $\big\lfloor 2 p\sqrt{\frac{M_0 |\mu^2|}{4  \lambda a}}\big\rfloor$, since we would like 
to have good integration accuracy in each unit-length interval inside of the parametric problem \eqref{param_inte} in the region
$\big[-p\sqrt{\frac{M_0 |\mu^2|}{4  \lambda a}},p\sqrt{\frac{M_0 |\mu^2|}{4  \lambda a}}\big]$. 
Note that in terms of the spacing parameter $a$, we have $m\sim O(a^{-\frac{3}{4}})$.
The remaining marginal tails outside of $\big[-p\sqrt{\frac{M_0 |\mu^2|}{4  \lambda a}},p\sqrt{\frac{M_0 |\mu^2|}{4  \lambda a}}\big]$ can be 
estimated with very few Gaussian points (or no points at all) if the relative contribution of the tail of the integrand 
to the total integral is negligible relative to our maximal target accuracy. 
For our experiments, we used Gauss-Hermite points (corresponding to a weight function 
of the form $e^{-cx^2}$, $c>0$) and  
Gaussian points generated from a weight function $e^{- \frac{4 \lambda a^3}{M_0^2}x_{i}^4}$, for the integration region 
$\big(-\infty,-p\sqrt{\frac{M_0 |\mu^2|}{4  \lambda a}}\big] \cup \big[p\sqrt{\frac{M_0 |\mu^2|}{4  \lambda a}},+\infty\big)$. By increasing 
the sample sizes in this region we were able to observe that the relative contribution of the integrand 
on $\mathbb{R}^d \setminus \big[-p\sqrt{\frac{M_0 |\mu^2|}{4  \lambda a}},p\sqrt{\frac{M_0 |\mu^2|}{4  \lambda a}}\big]^d$ 
to the total integrals on  $\mathbb{R}^d$ seems less than $10^{-12}$. Therefore we believe that the selected importance region
for integration can approximate the original problem with great relative accuracy (close to $10^{-11}$).\\

\begin{figure}[h!]
  \setlength{\columnsep}{-.55cm}
  \centering
  \begin{multicols}{2}
    \includegraphics[width=.57\textwidth, page=1]
    {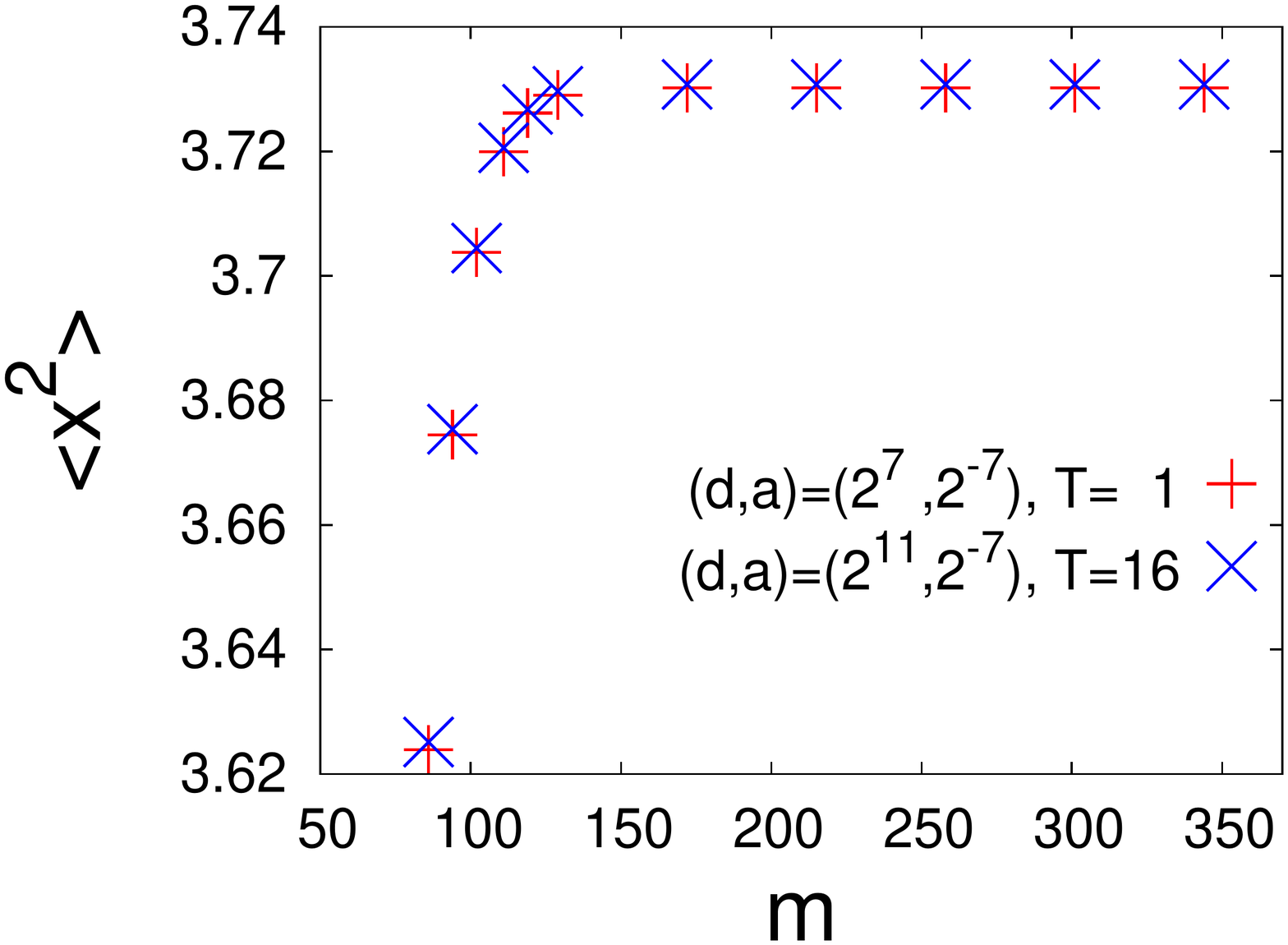}

    \includegraphics[width=.57\textwidth, page=2]
    {plots/gauss_quartic.pdf}
  \end{multicols}
  \vspace{-1cm}
  \begin{multicols}{2}
    \includegraphics[width=.57\textwidth, page=3]
    {plots/gauss_quartic.pdf}

    \includegraphics[width=.57\textwidth, page=4]
    {plots/gauss_quartic.pdf}
  \end{multicols}
  \vspace{-.5cm}
  \caption{Shown are the estimated values of $\langle x^2\rangle$
    (left) and $\langle x^4\rangle$ (right) in the anharmonic oscillator model obtained with recursive Gauss-Legendre quadrature 
  for the pairs $(d,a)=(2^{7},2^{-7})$, $(2^{11},2^{-7})$,
  $(2^{14},2^{-10})$ and $(2^{24},2^{-12})$, with corresponding time periods 
  $T=1.0$, $T=16$, $T=16$ and $T=4096$. In all cases the relative accuracy achieved with the corresponding maximal samples $m$ 
  seems to be more than 10 significant digits.}
  \label{fig:Anharm_osc_Gauss_conv}
\end{figure}

Figur \ref{fig:Anharm_osc_Gauss_conv} shows our calculations of the
two observables $\langle x^2\rangle$ and $\langle x^4\rangle$,
depending on the sample size $m$, for different values of lattice
points and spacing $(d, a)$.
Note that the highest $m\approx 4600$ was taken for $(d,a)=(2^{24},2^{-12})$, but 
even in this case the computations took no longer than 3 minutes with a standard PC.  For high values of $T$ (like $T=4096$) we 
observe very good results, as well, contrary to what has been observed with randomized QMC in \cite{Jansen:2012gj}.
We would like to remark that for the ground state energy $E_0$, which, by virtue of 
the virial theorem, is related to $\langle x^2 \rangle$ and $\langle
x^4 \rangle$ by $E_0 =  \mu^2  X^2 + 3 \lambda  X^4 + \frac{\mu^4}{16}$, 
the resulting estimates for $T=4096$ matches with the theoretical 
value in 5 significant digits, $E_0 = 3.8636669$, calculated in \cite{Blank79}, 
namely $\hat{E}_0 = 3.86367053759882$ for $(d,a)=(2^{24},2^{-12})$ and
$m\approx 4600$.
\clearpage

\section{Conclusions}

In this paper we have applied the methods of Quasi Monte Carlo and recursive
numerical integration to two quantum mechanical models discretized on a
Euclidean time lattice within the path integral approach.

The first model we have considered is a quantum mechanical rotor which 
is -- to some extent -- similar to higher dimensional spin systems 
such as non-linear $\sigma$-models.
For the quantum mechanical rotor 
we could, unfortunately, not find
a successful implementation of the QMC method to solve this model.
On the other hand, the method of recursive numerical integration led to
a much improved
accuracy even when compared to an optimal MCMC algorithm for which
we have chosen the cluster algorithm.
Conceptually, the error scaling of the recursive numerical integration
is at least exponentially fast in the number of integration points $m$.
In fig.~\ref{fig:gauss-dchi-m} we could indeed demonstrate that asymptotically 
this exponential error scaling is realized. Figure \ref{fig:dchi-t} shows that even 
for a small number of integration points the accuracy of the recursive integration 
method is already much higher than the one of the cluster algorithm with 
a correspondingly small number of samples. 

We remark that we have chosen the cluster algorithm as the MCMC method
since it avoids the increase of the autocorrelation time towards the 
continuum limit, a feature which is shared by the recursive numerical integration
technique by construction. 
The avoidance of autocorrelation times and our finding that a very high
accuracy can be reached already for a small number 
of integration points makes the recursive numerical integration technique
a very 
promising method for more difficult systems where 
only a small number of samples can be realized, e.g. in higher dimensions. 

The other model we looked at is the anharmonic quantum mechanical oscillator.
Here, we had found earlier that the QMC method shows an improved error
scaling \cite{Jansen:2012gj,Jansen:2013jpa,Ammon:2013yka}.
When the extent of the time lattice is kept short, $T \lesssim 1.5$, both, QMC and
recursive numerical integration show a comparable performance 
with an improved error scaling               
which is faster than $1/\sqrt{N}$.                                
However, as noted in \cite{Jansen:2013jpa} QMC becomes inefficient when the time
extent $T$ is made larger than $T=1.5$, independent of the value of the 
lattice spacing $a$. 

Therefore, we tested the performance of the recursive numerical integration method
for various choices of the dimension $d$ and lattice 
spacing $a$. As fig.~\ref{fig:Anharm_osc_Gauss_conv} shows, 
we can choose a very broad range for $(d,a)$ where we still 
find an extremely good performance of the recursive numerical integration method. In fact, the values 
of the time extent, $T=da$ can assume very large values such as $T=4096$  
and the values of $a$ can become tiny, e.g., $a=2^{-12}$, while still a
rapid convergence of the considered quantities is observed.   

In conclusion, we have tested two methods to evaluate the path-integral in
Euclidean time for quantum mechanical systems. These are the QMC and the
recursive numerical integration methods. While we could not find a successful
implementation for QMC in the case of the quantum mechanical rotor,
the technique of recursive numerical integration has been highly successful
with an exponentially fast error scaling. 
It will be very interesting to test this method for more complicated
1-dimensional models and, of course, for systems in higher dimensions.

\section*{Acknowledgment}
The authors wish to express their gratitude to Prof. Andreas Griewank
(Humboldt-Universit\"at zu Berlin) as well as Prof. Michael
M\"uller-Preussker (Humboldt-Universit\"at zu Berlin) for inspiring comments and
conversations, which helped to develop the work in this article.
H.L., J.V. and K.J. acknowledge financial support by the DFG-funded corroborative
research center SFB/TR9 and the DFG projects JA~674/6-1 and GR~705/13.


\bibliography{Article.bib}

\end{document}